\documentclass{article}

\usepackage[final]{neurips_2022_ml4ps}

\usepackage{mathtools}

\usepackage[utf8]{inputenc} 
\usepackage[T1]{fontenc}    
\usepackage{hyperref}       
\usepackage{url}            
\usepackage{booktabs}       
\usepackage{amsfonts}       
\usepackage{nicefrac}       
\usepackage{microtype}      
\usepackage{xcolor}         
\usepackage[T1]{fontenc}    
\usepackage{hyperref}       
\usepackage{url} 
\usepackage{booktabs}       
\usepackage{amsfonts}       
\usepackage{nicefrac}       
\usepackage{microtype}      
\usepackage{lipsum}		
\usepackage{graphicx}
\usepackage{natbib}
\usepackage{doi}
\usepackage{amsmath}
\usepackage{graphicx}
\usepackage{caption}
\usepackage{subcaption}
\usepackage{multirow}
\usepackage{float}
\usepackage{caption}
\usepackage{subcaption}

\setcitestyle{numbers,square}

\title{Pay Attention to Mean-Fields during Particle Cloud Generation}

\author{
Benno Käch\\
Deutsches Elektronen-Synchrotron DESY\\  Germany\\
\texttt{benno.kaech@desy.de} \\
\And
Isabell-A. Melzer-Pellmann \\
Deutsches Elektronen-Synchrotron DESY\\  Germany\\
\texttt{isabell.melzer@desy.de} \\
}

\begin{document}
\maketitle

\begin{abstract}
   The generation of collider data using machine learning has emerged as a prominent research topic in particle physics due to the increasing computational challenges associated with traditional Monte Carlo simulation methods, particularly for future colliders with higher luminosity. Although generating particle clouds is analogous to generating point clouds, accurately modelling the complex correlations between the particles presents a considerable challenge. Additionally, variable particle cloud sizes further exacerbate these difficulties, necessitating more sophisticated models. In this work, we propose a novel model that utilizes an attention-based aggregation mechanism to address these challenges. The model is trained in an adversarial training paradigm, ensuring that both the generator and critic exhibit permutation equivariance/invariance with respect to their input. A novel feature matching loss in the critic is introduced to stabilize the training. The proposed model performs competitively to the state-of-art whilst having significantly fewer parameters.\footnote{The model and the code to reproduce the results are available under \url{https://github.com/kaechb/MDMA}}

\end{abstract}

\section{Introduction}
Machine Learning (ML) has been applied in High Energy Physics (HEP) for decades. With the recent advances in ML, multiple applications of it have become the new standard in all stages of data analysis in HEP~\cite{mlinhep}. Although most of them were in a supervised training frame, unsupervised approaches also gain interest. In HEP, detailed simulations of the physical processes, which almost perfectly describe the details of the experimental measurement, are commonly available. These Monte Carlo simulations (MC) provide labelled data and are needed in large numbers to cover all areas of the physical phase space. The simulations for the CMS detector at the Large Hadron Collider, for example, require about $50 \%$~\cite{mccms} of the current CMS computing budget. An even more significant number of simulations will be needed for the coming high-luminosity phase of the LHC~\cite{hllhc}. The recent progress in generative models sparked interest in the HEP community, as an enhancement, and/or partial replacement, of the sequential approach of MC with generative models, which can be efficiently parallelized on GPUs. 
This study investigates the generation of PYTHIA~\cite{pythia} jets using a Generative Adversarial Network (GAN) \cite{GAN}. The publicly available \textsc{JetNet} dataset, suitable metrics for it and other existing models are briefly introduced in Section~\ref{sec:dataset}. In the following, the model and its training paradigm are presented in Section~\ref{sec:architecture}, and its results are compared with the state-of-art (SoA) in Section~\ref{sec:results}.

\section{Related Work}
\label{sec:dataset}
Raghav et al. introduced the \textsc{JetNet} datasets~\cite{mpgan}, and at the same time, proposed a suitable model to generate such jets. The dataset contains $\sim 180,000$ samples, split into 70\% training, 15\% validation, and 15\% test data.
There are samples for quark- as well as boson-initiated jets with an energy of about 1~TeV, all generated using PYTHIA 8.212~\cite{pythia}. These are available for light-quark, top-quark, gluon and W/Z boson intitiated jets. The jet constituents are considered to be massless and can therefore be described by their 3-momenta, or equivalently by their transverse momentum $p_\textrm{T}$, pseudorapidity $\eta$, and azimuthal angle $\phi$. In the \textsc{JetNet} dataset, these variables are given relative to the jet axis: 
$p_{\textrm{T},i}^{\textrm{rel}} \coloneqq p_{\textrm{T}.i}^{\textrm{particle}}\!/p_\textrm{T}^{\textrm{jet}}$,
$\,\eta_i^{\textrm{rel}} \coloneqq \eta_i^{\textrm{particle}}\!-\eta^{\textrm{jet}}$, and 
$\,\phi_i^{\textrm{rel}} \coloneqq (\phi_i^{\textrm{particle}}\!-\phi^{\textrm{jet}})\bmod 2\pi$, 
where $i$ runs over the particles in a jet. Each jet is constrained to contain at most 150 particles.  Most available models to date train on a reduced dataset, which is constrained to the hardest $30$ particles per jet. The restriction to 30 (and also 150) particles leads to the jets not being centered, meaning that the original jet axis including all particles from PYTHIA is not equal to the one calculated from the reduced number of constituents. To overcome this, the authors of~\cite{epic} use an additional preprocessing step and center the jets. On the JetNet-150 dataset, the effect is not expected to be very strong, as only a negligible amount of the (original) jets contains more than 150 constituents (also these particles are usually very soft).

A jet's invariant mass $m_{\textrm{jet}}$ is an essential high-level feature containing important physics information. It is a global variable that depends on the correlations between the jet constituents and provides an essential one-dimensional feature to measure the performance of the generative model. For the relative quantities above, the scaled jet mass can be defined as $(m^{\textrm{rel}})^2=(\sum_i E^{~\textrm{rel}}_i)^2-(\sum_i \vec{p}^{~\textrm{rel}}_i)^2 = m_{\textrm{jet}}^2/p_{\textrm{T,jet}}^2$. 
An additional set of features is constructed in~\cite{mpgan} by using the Energy Flow Polynomials (EFP)~\cite{efp}, which form a complete set of jet substructure observables. 

To quantify the performance of the models on this reduced set of features, the use of a Wasserstein-1 distance on one-dimensional summary statistics was first proposed in~\cite{mpgan}, where a model based on message-passing GAN~\cite{messagepassingNN} is presented as well. Their generated data on the reduced dataset (where only the 30 hardest particles are modelled) is almost compatible with in-sample distances. This is also achieved by several other models, ranging from Normalising Flows\cite{NFbenno,transflow} and GANs~\cite{epic} over to diffusion models~\cite{diffusionjetnet,diffusionjetnet2}. These models can be grouped into two classes: models using conditioning only on the number of particles in a jet which is to be generated, and models that rely on additional jet observables as input.  In Raghav et al.'s initial publication~\cite{mpgan} they have shown that learning the distribution of the number of particles in a jet is extremely challenging. Thus, we refer to models using only the  number particles in a jet as "unconditioned" models and to models using additional jet observables as "conditioned" models - although this might seem a bit unintuitive at first. In previous work~\cite{NFbenno}, we found that unconditioned discrete Normalising Flows are unable to model correlations in the data correctly, but they perform quite well when  the mass is added as an additional input to the model. However, this is more patchwork than a viable solution: when finding that the model is not able to model the mass correctly, we supply the mass as input to the proposed model. Intuitively, this makes modelling the mass significantly easier but by supplying a jet observable to the model, one also takes a powerful performance test away. Thus  the conditioning on the mass is replaced in a following publication~\cite{transflow} by using an unconditioned Normalising Flow as a prior for a GAN, which is based on a transformer architecture. 

In the following, the proposed model's performance is only compared to unconditioned models. The remaining models~\cite{transflow,epic,mpgan,jetneteval} can again be classified into two groups based on their computational scaling with the number of particles $n$. Suppose the model allows a direct exchange of information between individual particles. In that case, the model will scale with $\mathcal O(n^2)$~\cite{transflow,mpgan} which makes the training infeasible when moving to a higher particle multiplicity. 
Raghav et al.~\cite{jetneteval} provide an in-depth study on different metrics to evaluate the performance of a model, showing that the Wasserstein-1 distance is biased. Therefore, they propose two additional metrics: the kernel and Frechet physics distances (KPD/FPD).  In the same publication they were also the first to propose a model that scales linearly on the number of particles, based on the set transformer architecture~\cite{settransformer}. However, they only show their model performance on the reduced dataset in this publication. Buhmann et al.~\cite{epic} were the first to present a model scaling linearly on the \textsc{JetNet}150 gluon and quark datasets and are as such the only ones that the proposed model can be benchmarked against.


\section{Architecture}
\label{sec:architecture}
The model proposed here is inspired by our previous model~\cite{transflow}, which also relies on attention, but uses an unconditioned normalising flow as a prior, whose output is then passed through a transformer-inspired post-processing network that is trained adversarially. However, the Normalising Flow leads to a considerable overhead for computations and model size. Also, calculating attention naively between every particle to exchange information between particles leads to a scaling that is quadratic in the number of particles, which is not feasible as we move to higher number of particles to be modelled. 
The main challenges that are thus tackled in this study are to replace the normalising flow prior and find an information aggregation mechanism that scales linearly with the particle multiplicity.  

Thus, the final model setup consists only of a critic and a generator, as typically used for GANs. Both networks take a variable number of particles and are built to be equivariant under the permutation of the input. The architecture also allows the masking of particles that subsequently have no influence on the model's output - which becomes more and more important at a higher variance in particle multiplicities as otherwise the additional for-loop over the particle multiplicity distribution in~\cite{epic} cannot be easily parallelized on a GPU.
To aggregate information from the different particles, a BERT~\cite{BERT}-inspired classification token is used. In this context however, it is illustrative to think of this token as a mean-field particle, which is created by the particles, and the interaction of all particles with each other is approximated by the interaction with this mean-field particle only. The mean-field particle is initialized by the sum of all particles normalized by some constant. We do not initialize it as the mean (as the name suggests) because of the training process. As  similar-sized jets are bucketed to batches and padded only to the largest particle multiplicity per batch for more GPU efficiency, the batches do not have the same number of particles per batch, and using the mean would be biased. The operation that is used to aggregate information from all particles is cross-attention~\cite{attention} between a high-dimensional representation of the mean-field particle and a high-dimensional representation of the other particles. The attention-based aggregation allows the critic/generator to select which particles are important to distinguish a real (simulated) jet from a fake (generated) dynamically, and disregard unimportant particles. We expect this to become more and more relevant for higher particle multiplicities and the presence of large variances in the energy of the individual particles. 

\begin{figure}
    \centering
    \includegraphics{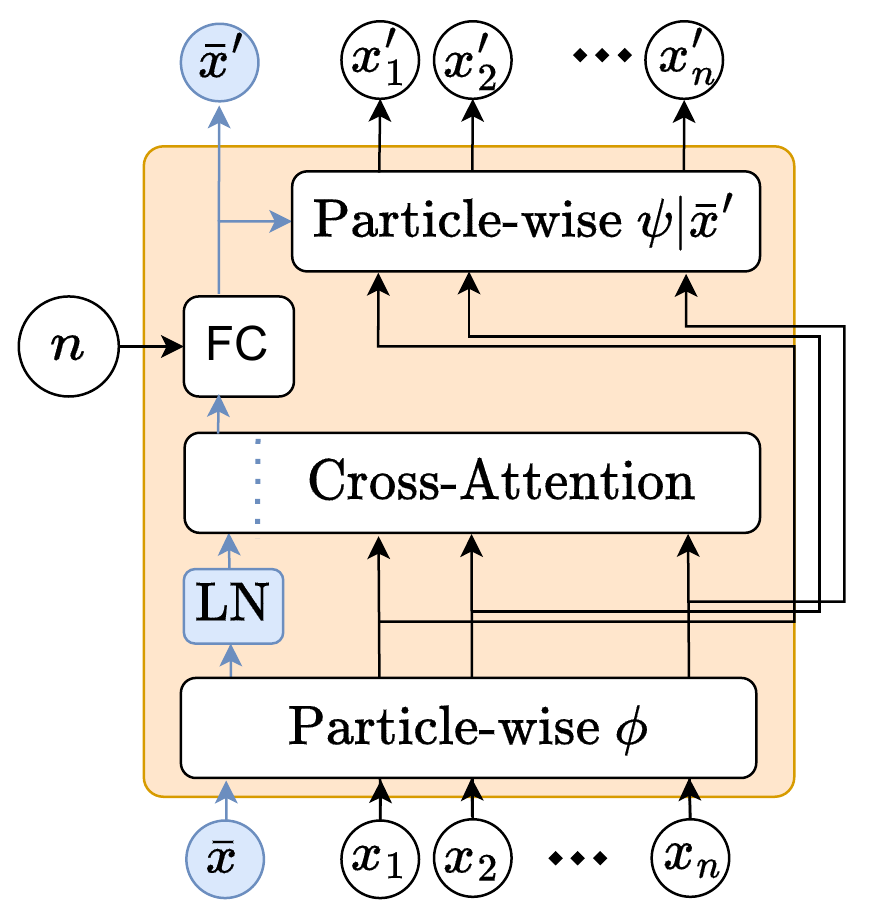}
    
    \caption{Minimal building block used in the architecture. The input particles and the mean-field particle are first passed through the same particle-wise linear layer and activation to be mapped to a higher dimension $h$ where the information exchange between the particles and mean-field takes place via cross-attention. The mean-field particle is passed through a Layer Norm (LN) \cite{Layernorm} before this aggregation. The updated mean-field is then concatenated with the particle multiplicity in the jet and passed through a fully-connected (FC) layer which maps it down to the initial dimension $l$. This update mean-field is not only the output of the building block, but is also used to update the particles by applying another particle-wise layer using the mean-field as a condition and mapping the particle down to their initial dimension $l$.}\label{fig:critic}
    
\end{figure}

The minimal building block for the critic/generator is shown in Fig~\ref{fig:critic}. Multiple blocks are stacked to build the main body of the architecture. However, before the first block, a particle-wise linear map is applied to the particles, mapping them to an $l$-dimensional latent space. Although critic and generator use the same building blocks, there are notable differences. They differ in the residual connections that are not shown in Fig.~\ref{fig:critic},  the activations used, the layers coming after the main body, and the use of weight-normalized linear layers in the critic specially designed for GANs~\cite{weightnorm}. For the generator, there are residual connections between particles entering and particles coming out of one block, whereas for the critic there is instead a connection between the entering and outgoing mean-field particle. LeakyReLU activations with a slope of $0.01$ are applied in the critic, while  Gaussian Error Linear Units~\cite{GELU} are used in the generator. After the last block, another particle-wise linear layer is applied in the generator architecture to map from the latent dimension $l$ down to the initial three dimensions per particle. Similarly for the critic, a 2-layer neural network with a linear activation is applied after the last layer to the mean-field particle to obtain the score ("realness") of a generated or real jet. Note that the critic is also able to change the representation of the particles, because of the exchange of information between the mean field particle and the particles in the critic architecture. This allows the critic to aggregate information from the indirect interaction of particles (as all particles interact with the mean-field particle, which then interacts with the particles). 

Different ways of transferring information from the mean-field to the particles were tested. The simplest approach is pure addition of the mean-field to all other particles. Although this works to a certain extent, problems arise as this aligns all particles with the mean-field particle. This impacts the calculation of cross-attention weights, as those measure the similarity between the mean-field and its inputs in some latent space. A particle-wise layer, which takes the particles and the mean-field as input, comes at the cost of introducing more parameters, but its introduction is necessary for a competitive performance. 
Supplying the number of particles in a jet  explicitely to the model is crucial for the performance of the generator and critic. The model is conditioned on the number of particles in a jet $n$ implicitly by choosing the dimension of the input noise as $n$ (number features per particle). But since attention is a weighted sum, where the weights add up to unity, the mean-field particle is by construction agnostic to the number of elements during an aggregation step. We found that it is enough to include the number of particles of a given jet as an additional input while mapping down the mean-field particle to its embeddeding dimension. Aggregating the information from this input with a gated linear unit (GLU)~\cite{glu} was also tested, but this gave worse results (and notably uses the same number as just using the number of particles as an additional input for the linear layer). The generator does not use additional degrees of freedom for a global state, resulting in the latent space of the generator having the same dimension as the space of the data that is modelled. 

\subsection{Training and Evaluation}
Different training procedures are tested and compared, including the Vanilla GAN~\cite{GAN} training, an LSGAN~\cite{LSGAN} and a WGAN training with gradient penalty~\cite{WGAN,wgangp}. Optimization is done using Adam~\cite{adam} with a weight loss coefficient of 0.01 for the critic, and the learning rate is varied with Cosine Annealing after a linear warmup to $10^{-4}$ that is commonly used with transformers~\cite{attention}. The most stable results were obtained with the LSGAN training.  

The models were trained for 3000 Epochs. The generator has $\sim87'000$ parameters, whereas the critic has $\sim43'000$ which is significantly smaller than the $425'000/313'000$ used in \cite{epic}. For generation, the model needs $\sim 9.2\ \mu s$ per jet, when generating a sample of $500'000$ events on a NVIDIA A100 GPU with a batchsize of $75'000$. 

In Ref.~\cite{epic}, ten times as many samples as available in the testing sample are generated, and the mean and standard deviation is estimated over ten runs, sampling without repetition. As the EPiC-GAN model (introduced in Ref.~\cite{epic}) is the only unconditional model that is currently available, the same evaluation procedure is adopted here to simplify comparison. The models used for the evaluation are chosen based on the best $W_1^m$ distance, albeit checkpoints with underfluctuations of the Wasserstein distance are vetoed using the FPD.
To allow a fair comparison between the generalization of the proposed model to EPiC-GAN, the same hyperparameters are used for all datasets - although it is worth mentioning that on the light quark dataset a bigger network size led to slightly better performance. For the final comparison, the model's hyperparameters were first optimized on the top-quark dataset, as the two-peak structure is expected to make this dataset the most difficult one to model. Unfortunately the authors of~\cite{epic} do not use the train/valid/test split offered by the \textsc{JetNet} library, making a direct comparison difficult, as most likely some events of the testing set used to quantify the proposed model's performance have been seen by their model during training. Still, both models use about 70\% of the data for training, 15\% for validation and 15\% for testing. As the Wasserstein distance is biased with a bias proportional to the size of the sample being tested~\cite{jetneteval}, it is important to use the same sample size during evaluation. 

For the calculation of the $W_1^P$ and $W_1^{EFP}$ the variance weighted mean is quoted. The weighted mean is calculated from the variance of the metric on the different variables it is calculated on.

\subsection{Mean-field matching}
To accelerate the generator training progress, a mean-field matching loss is introduced, where the generator additionally optimizes an L2 norm of the final mean-field of the critic between real and fake jets. Note that this is similar to feature matching proposed by Goodfellow et al. in~\cite{gantricks}, but they propose to minimize the squared difference of the mean. Since there is no one-to-one correspondence between real and fake jets, the loss reaches its minimum at the standard deviation of the mean-field for real jets. Our studies show that this generally increases training convergence time, however the loss sometimes diverges after reaching the minimum. Thus the model is only trained with feature matching for 30k gradient steps. 

\section{Results}\label{sec:results}
To summarize the most important aspects of the model it is referred to as a Mean-fielD Matching Attentive (MDMA) GAN in the following.
In Table~\ref{tab:results} the results are compared to the EPiC-GAN model, where results for it are available. The in-sample metrics (IN) calculated between training and testing sets are given for comparison as well. To obtain the results for the EPiC-GAN, the checkpoint provided in their repository was used. We also apply their additional centering step to their generated data (and also the true data when calculating the metric with respect to their dataset), but we would like to mention that the irreversible preprocessing step applied to the training sample impedes the benchmarking. As there is no available checkpoint on the W/Z sample, the model performance can only be compared to in-sample distances. The MDMA-GAN model performs competitively to EPiC-GAN, the metrics compared to their SoA model are in general compatible or better. Some notable differences are visible on $W_1^P$, where the model consistently performs significantly better, albeit the best checkpoint was chosen on $W_1^m$.  

\begin{figure}
  \centering
  \includegraphics[width=\textwidth]{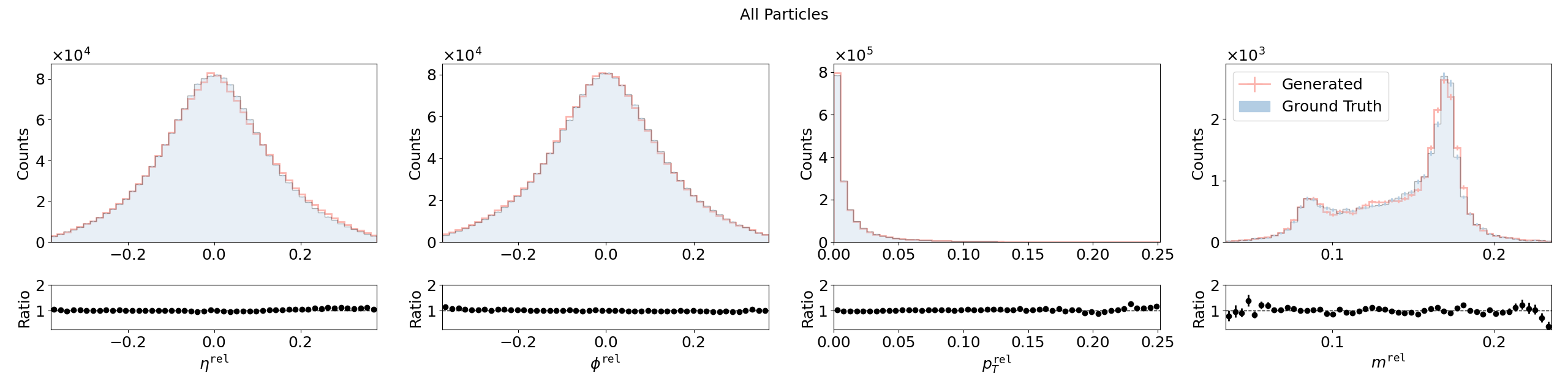}
  \caption{Distribution of all particles $(\eta^{rel},\phi^{rel},p_T^{rel})$ and the invariant mass. The generated data is compared to a hold out data set, and a ratio is shown below the plot.}
  \label{fig:results}
\end{figure}

We propose to not only select the best checkpoint based on the $W_1^M$ metric, but also based on the FPD metric - as it is also most sensitive in this case study confirming the conclusion drawn by Kansal et al.~\cite{jetneteval}. However, note that Raghav et al. recommend a sample size of at least 50'000 jets as the extrapolation used in the metric is otherwise not stable. But in order to be able to fairly compare to EPiC-GAN (where samples of the size of 25'000 are used),  only $\sim25'000 $ samples are used here for testing as well. The KPD metric saturates fairly quickly, as for the proposed model the KPD is always compatible with 0. In Fig.~\ref{fig:results} a visualization of the results are shown only for the top-quark dataset, as usually it is quite difficult to extract information of the model from these simplified 1-dimensional plots, and the Wasserstein-1 distances stand in a one-to-one correspondence with them.
\begin{table}[h]

\caption{Comparison between the best performing models from Ref.~\cite{mpgan} (EPiC), with the model proposed in this study (MDMA) on all datasets. We highlight scores where one model performs significantly better than the other. For reference, the in-sample distance between the training and testing sample (IN) are also given. As there are no models available for the W/Z sample for the EPiC-GAN, only results of the proposed model together with the in-sample distances are quoted. 
}
\label{tab:results}

    \centering
    \vspace{0.2cm}
    \resizebox{1\textwidth}{!}{
    \begin{tabular}{ll|lllll}Jet Class & Model & $W_1^M (\times 10^{3})$ & $W_1^P
(\times 10^{3})$ & $W_1^{EFP}(\times 10^{5})$ &$KPD (\times 10^{4})$&$FPD(\times 10^{4})$\\\cline{1-7}

\cline{1-7}\multirow{2}{*}{Light Quark} & 

EPiC & $\mathbf{0.5 \pm 0.1}$ & $1.25 \pm 0.09$ & $0.81 \pm 0.16$ & $-0.0 \pm 0.1$ & $5 \pm 1$ \\&
MDMA & $0.7 \pm 0.1$ & $\mathbf{0.37 \pm 0.05}$ & $0.82 \pm 0.14$ & $-0.05 \pm 0.07$ & $\mathbf{3.8 \pm 0.8}$ \\&
IN & $0.5 \pm 0.1$ & $0.15 \pm 0.04$ & $0.59 \pm 0.16$ & $-0.0 \pm 0.2$ & $3 \pm 2$ \\\cline{1-7}
\multirow{2}{*}{Gluon} & 

EPiC & $0.5 \pm 0.1$ & $1.13 \pm 0.04$ & $\mathbf{0.93 \pm 0.14}$ & $0.06 \pm 0.05$ & $4.2 \pm 0.7$ \\&
MDMA & $0.5 \pm 0.1$ & $\mathbf{0.27 \pm 0.02}$ & $1.12 \pm 0.13$ & $\mathbf{-0.05 \pm 0.03}$ & $\mathbf{1.4 \pm 0.5}$ \\&
IN & $0.6 \pm 0.2$ & $0.048 \pm 0.009$ & $0.79 \pm 0.24$ & $-0.1 \pm 0.1$ & $3.2 \pm 0.8$ \\\cline{1-7}
\multirow{2}{*}{Top Quark} & 

EPiC & $0.69 \pm 0.08$ & $0.65 \pm 0.03$ & $2.67 \pm 0.39$ & $1.7 \pm 1.0$ & $22 \pm 1$ \\&
MDMA & $\mathbf{0.57 \pm 0.09}$ & $\mathbf{0.10 \pm 0.02}$ & $\mathbf{2.12 \pm 0.64}$ & $\mathbf{-0.0 \pm 0.2}$ & $\mathbf{5.3 \pm 0.9}$ \\&
IN & $0.42 \pm 0.09$ & $0.12 \pm 0.04$ & $1.22 \pm 0.32$ & $-0.1 \pm 0.2$ & $1.2 \pm 0.6$ \\\cline{1-7}
\multirow{1}{*}{W} & 

MDMA & $\mathbf{0.31 \pm 0.02}$ & $\mathbf{0.20 \pm 0.04}$ & $\mathbf{0.284 \pm 0.040}$ & $\mathbf{0.01 \pm 0.02}$ & $\mathbf{1 \pm 4}$ \\&
IN & $0.18 \pm 0.05$ & $0.14 \pm 0.04$ & $0.136 \pm 0.021$ & $-0.00 \pm 0.01$ & $1 \pm 1$ \\\cline{1-7}
\multirow{1}{*}{Z} & 

MDMA & $\mathbf{0.27 \pm 0.03}$ & $\mathbf{0.23 \pm 0.04}$ & $\mathbf{0.373 \pm 0.069}$ & $\mathbf{-0.01 \pm 0.02}$ & $\mathbf{1 \pm 4}$ \\&
IN & $0.18 \pm 0.03$ & $0.24 \pm 0.07$ & $0.152 \pm 0.042$ & $0.00 \pm 0.02$ & $2 \pm 1$ \\\cline{1-7}
  \end{tabular}}

\end{table}

\section{Summary}
In this study a particle cloud generation model is presented, which scales linearly with the number of particles/tokens due to the use of cross-attention to an artificial mean-field particle. The proposed model performs competitively to the SoA at time of publication using significantly fewer parameters. Differences between in-sample distances and generated samples are still observed. But as the evaluation metrics used heavily depend on the sample size, and ML models generally profit from increased sample size, it remains interesting for future work whether an increase in data would resolve the remaining discrepancies. 
It is important to note that particle clouds are similar to point clouds, although possibly even more complex due to the existance of complex correlation between the particles as for example the mass. 
Thus the use of this model in ML applications working with point clouds seems promising.
For further work the model will be applied to datasets with more particles and its scaleability to problems with an even higher dimensionality as calorimeter shower generation will be evaluated.

\section{Acknowledgements}

Benno K\"ach is funded by Helmholtz Association’s Initiative and Networking Fund through Helmholtz AI (grant number: ZT-I-PF-5-64).
This research was supported in part through the Maxwell computational resources operated at Deutsches Elektronen-Synchrotron DESY (Hamburg, Germany).
The authors acknowledge support from Deutsches Elektronen-Synchrotron DESY (Hamburg, Germany), a member of the Helmholtz Association HGF.


\newpage
\newpage
\bibliographystyle{unsrtnat}


\bibliography{ref}  

\newpage




\end{document}